\documentclass[11pt]{article}
\usepackage{moriond,epsfig,amssymb}

\newcommand{\fig}[1]{~\ref{fig:#1}}

\newcommand{\TeV}{\,{\rm TeV}}
\newcommand{\diag}{\hbox{diag}\,}
\newcommand{\MeV}{\,\hbox{\rm MeV}}
\newcommand{\eV}{\,\hbox{\rm eV}}
\newcommand{\he}{^4 \textrm{He}}
\newcommand{\nnu}{N_\nu}

\newcommand{\nue}{\nu_e}
\newcommand{\nueb}{\bar\nu_e}

\newcommand{\nus}{\nu_{\rm s}}

\def\be{\begin{equation}}
\def\ee{\end{equation}}
\def\bea{\begin{eqnarray}}
\def\eea{\end{eqnarray}}

\begin{document}
\vspace*{4cm}
\title{STERILE NEUTRINOS: FROM COSMOLOGY TO EXPERIMENTS}

\author{GUIDO MARANDELLA}

\address{Scuola Normale Superiore and INFN \\ Piazza dei Cavalieri 7, Pisa, I-56126,
Italy}

\maketitle\abstracts{We analyze the oscillation signals generated by
  one extra sterile neutrino. We fully take into account the effects
  of the established oscillations among active neutrinos. We
  analyze the effects in solar, atmospheric, reactor and beam experiments,
  cosmology and supernov\ae. We analyze the impact of the LSND anomaly
  on cosmology showing, in the full 4 neutrino context, that it is not
  compatible with standard cosmology. We identify the still allowed regions of
  the parameter space, outlining which are the future experiments
  which can improve the bounds.}

\section{Introduction}

The solar and atmospheric neutrino anomalies are today very well
explained by oscillations among the three active Standard Model (SM)
neutrinos. The contribution from a fourth sterile neutrino once was an
alternative explanation the the observed anomalies, but today it can
represent only a subleading contribution to the standard scenario of
active-only oscillations. Many extensions of the Standard Model
foresee the existence of fermions which might have a mass of the order
$\TeV^2/M_{{\rm Pl}}$. A few candidates are, in alphabetic order,
axino, branino, dilatino, familino, Goldstino, Majorino, modulino,
radino.
The relevant questions today are then the following: how large can be a
subdominant sterile neutrino effect and where do we have to look for
it? Of course the mixing of a sterile neutrino with the active ones
affects directly the neutrino experiments (solar, atmospheric, reactor
and beam). However neutral fermions with $\eV$-scale mass, which is
the definition of ``sterile neutrinos'' for us, are typically stable
enough to give effects in cosmology (Big-Bang Nucleosynthesis, Cosmic
Microwave Background, Large Scale Structure). Furthermore
active-sterile neutrinos mixing would have modified the fluxes of active neutrinos coming from the SN1987A supernova with respect to the standard scenario (with no sterile neutrinos). 

After presenting our non-standard parametrization of active-sterile
mixing (Sec. \ref{sec:parametrization}) we will summarize the
effects of a sterile neutrino on
solar and KamLAND experiments (Sec. \ref{sec:solar}),  on atmospheric, reactor neutrinos, short
and long-baseline neutrino beams (Sec. \ref{sec:atm}), on cosmology
(Sec. \ref{sec:cosmo}) and on supernov\ae{}
(Sec. \ref{sec:SN}). The results
will be shown in a final set of plots which will identify the regions
of parameter space which are excluded (in the case of experiments) or
strongly disfavored (in the case of cosmology and supernov\ae). To
avoid to show an unreadable plot, we show only a few lines and
shade the whole region excluded by both solar and atmospheric (and
reactor and beam) experiments. To better understand the separated
bounds we refer to the plots of \cite{Cirelli:2004cz}.

\section{Parametrization of active-sterile mixing}
\label{sec:parametrization}

Active-sterile neutrinos mixing is usually considered assuming that
the initial active neutrino $|\nu_{\rm a}\rangle$ oscillates with a single
large $\Delta m^2$ into a mixed neutrino
$\cos\theta_{\rm s}|\nu'_{\rm a}\rangle + \sin\theta_{\rm s}|\nu_{\rm
  s}\rangle$. We relax these simplifying assumptions and study the more general $4$-neutrino context. 

\begin{center}
    \begin{figure}
     $$\includegraphics[width=4.5cm]{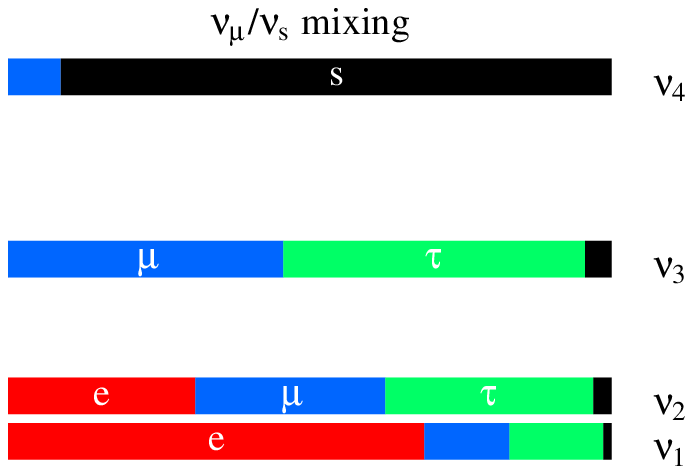} \hspace{2cm}
     \includegraphics[width=4.5cm]{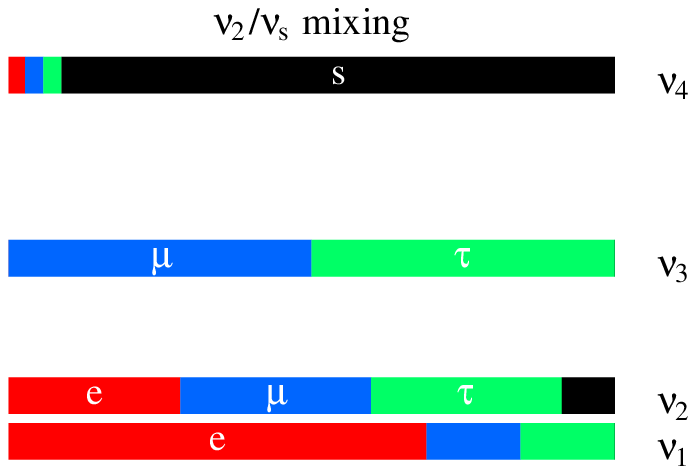}$$
     \caption{ \label{fig:spettrias} {\bf Basic kinds of four neutrino mass spectra}. 
    Left: sterile mixing with a flavor eigenstate ($\nu_\mu$ in the picture).
    Right: sterile mixing with a mass eigenstate ($\nu_2$ in the picture).}
    \end{figure}  
  \end{center}

In absence of sterile neutrinos, we denote by
$U$  the usual $3\times 3$ mixing matrix that relates
neutrino flavor eigenstates $\nu_{e,\mu,\tau}$ to active
neutrino mass eigenstates $\nu^{\rm a}_{1,2,3}$
as $\nu_\ell = U_{\ell i} \nu_i^{\rm a}$
($i=\{1,2,3\}$, $\ell = \{e,\mu,\tau\}$). In order to parametrize the mixing of an additional sterile neutrino we introduce  a complex unit 3-versor $\vec{n}$ which defines the combination of active neutrinos
\begin{equation}
  \vec{n}\cdot \vec{\nu} = n_e \nu_e + n_\mu \nu_\mu + n_\tau \nu_\tau = 
n_1\nu_1^{\rm e} + n_2 \nu_2^{\rm a} + n_3 \nu_3^{\rm a}\qquad (n_i = U_{\ell i}n_\ell)
\end{equation}
to which the sterile neutrino mixes with an angle $\theta_{\rm s}$. With this parameterization the fourth mass eigenstate is a superposition of the sterile neutrino neutrino $\nus$ and of the combination $\vec{n}\cdot \vec{\nu}$: $\nu_4 =  \nu_{\rm s} ~\cos\theta_{\rm s} + n_\ell \nu_\ell ~\sin\theta_{\rm s}$. The connection between our parametrization and the usual ones can be found in \cite{Cirelli:2004cz}.
In Fig. \fig{spettrias} we exemplify the two limiting cases we focus
on: mixing with a flavor eigenstate (fig.\fig{spettrias}a) and mixing
with a mass eigenstate (fig.\fig{spettrias}b).

From now on we will consider a $3+1$ scheme, with normal hierarchy and
we will set $\theta_{13}=0$. We have checked that allowing
$\theta_{13}$ to assume its maximum allowed value leads to at most
minor corrections to our analysis.

\section{Sterile neutrinos and solar (and KamLAND) experiments}
\label{sec:solar}

The solar neutrino anomaly is today explained by $\nue \to \nu_{\mu,
  \tau}$ oscillations in the LMA-MSW region with mixing parameters
$\tan^2 \theta_{{\rm sun}}  = 0.41 \pm 0.05, \; \; \Delta m^2_{{\rm
    sun}}  = (7.1 \pm 0.6) \cdot 10^{-5} \eV^2$. We looked for a
subdominant sterile neutrino signal in the experimental data. The work
we have done is the following.
First of all we compute, in the full 4 neutrinos context, the survival
probability $P_{ee}$ at the detection point for an electron neutrino
produced inside the sun. The most complicated task is to follow the
evolution of the neutrinos inside the sun, which is better achieved in terms of the $4
\times 4$ neutrino density matrix. Possible
non-adiabaticities are considered with usual crossing probabilities
$P_C$ which however are computed analytically in a non-standard way
\cite{Cirelli:2004cz}. This allows to develop a faster numerical code.
Then we fit all the experimental data: the SNO CC+NC+ES
  spectra; the total CC, NC and ES rates measured by SNO with
  enhanced NC sensitivity; the Super-Kamiokande ES
  spectra; the Gallium rate obtained averaging
the most recent  SAGE, Gallex and GNO data; the Chlorine rate; the
  KamLAND reactor anti-neutrino data. 
We calculate a $\chi^2$ as function of the 2 parameters that
describe sterile oscillations,
marginalizing the full $\chi^2$ with respect to
all other sources of uncertainty {\em including the LMA parameters}
$\Delta m^2_{\rm sun}$ and $\theta_{\rm sun}$. This is motivated by
the fact that experiments allow only relatively minor shifts from the LMA point.

No statistically significant evidence of a sterile neutrino has been found. The
results are shown in Fig. \ref{fig:all} where we shaded the region
excluded at $99\%$ C.L. Notice that in the same plot we shaded also
the region excluded at the same C.L. by atmospheric, reactor and beam
experiments (Sec. \ref{sec:atm}). For separated plots see
\cite{Cirelli:2004cz}.  In \cite{Cirelli:2004cz} we also
outlined a few promising signals. The most powerful one would be to
discriminate a 2 \% shift from the LMA region of the survival
probability $P_{ee}$ at sub-MeV energies. In fact it is at such
energies that LMA oscillations allow a $\nu_1$ component in the solar
neutrino flux: at higher energies matter effects are so strong that
the solar flux is constituted by $\nu_2$ only. The sterile neutrino
$\nu_s$ can mix with $\nu_1$ either directly mixing to it or,
independently on the mixing, because the two levels cross \cite{Cirelli:2004cz}.

\section{Sterile neutrinos in atmospheric, reactor and beam experiments}
\label{sec:atm}

The atmospheric anomaly is today explained by $\nu_\mu \to \nu_\tau$
oscillations with parameters $\sin^2 2 \theta_{{\rm atm}} = 1.00 \pm
0.05, \; \; \Delta m^2_{{\rm atm}} = (2.0 \pm 0.4) \cdot 10^{-3}
\eV^2$. We looked for subleading effects due to sterile neutrinos in
the data.
We followed the evolution of the $4 \times 4$ neutrino density matrix
$\rho$ from the production to the detection point. In each medium
(air, mantle, core) the evolution is given by a diagonal matrix of
phases $ \diag \exp (-2 i \delta)$ where $\delta_i =
m_{\nu_{mi}}^2L/4E_\nu$. We fitted all most recent results
and built a global $\chi^2$ which we marginalized with respect to
$\Delta m^2_{{\rm atm}}$ and $\theta_{{\rm atm}}$. This makes the
computation much simpler and is motivated by the fact that the
experimental data allow only minor shifts from the $\Delta m^2_{{\rm
    atm}}, \theta_{{\rm atm}}$ point. We do not include in our fits
the LSND data, which have been
  analyzed separately. In
  Sec. \ref{sec:cosmo} we analyze the cosmological impact of the LSND
  anomaly.

It is useful to distinguish two class of experiments.
In the data which are not sensitive to the atmospheric anomaly
  ({\sc Chooz}, {\sc Bugey}, {\sc CDHS}, {\sc CCFR}, {\sc Karmen},
  {\sc Nomad}, {\sc Chorus}) one has essentially to handle with vacuum
  oscillations. Disappearance experiments provide the dominant
  constraints. Instead, the experiments which see the atmospheric
  anomaly (SK, MACRO and K2K) probe sterile neutrinos in a significant
  but indirect way, essentially by the zenith-angle
spectra of $\mu$-like events with TeV-scale energies.

It is useful to compare the sensitivity of the these two classes of experiments.
Since there are no MSW resonances,
all these experiments are sensitive only to relatively large sterile mixing,
$\theta_{\rm s} \gtrsim 0.1$.
Sterile mixing with $\nu_e$ (and with the $\nu_1$ and $\nu_2$ mass eigenstates 
that contain a sizable $\nu_e$ fraction)
is better probed by reactor experiments,
although $e$-like events at SK
extend the sensitivity down to smaller values of $\Delta m^2$.
On the contrary atmospheric experiments give more stringent tests
of $\nu_{\rm s}/\nu_\tau$ mixing and of $\nu_{\rm s}/\nu_\mu$ mixing.

The results are shown in Fig. \ref{fig:all} where we shaded the
region excluded at $99\%$ C.L. (as already said above, for
separated exclusion plots for solar and atmospheric experiments see  \cite{Cirelli:2004cz}).

\section{Sterile neutrinos and cosmology}
\label{sec:cosmo}

Cosmology can test sterile neutrinos in the following ways.

{\bf Big Bang Nucleosynthesis (BBN).} BBN probes the total energy
density of the Universe at $T \sim (0.1 \div 1)\MeV$. Given a few
input parameters 
(the effective number of thermalized relativistic species,
the baryon asymmetry $n_B/n_\gamma=\eta$, 
and possibly the $\nu_\ell /\bar\nu_\ell$ lepton asymmetries)
BBN successfully predicts the abundances of several light
nuclei. Today $\eta$ is best determined within minimal cosmology by
CMB data to be \cite{Spergel:2003cb} $\eta = (6.15 \pm 0.25) \cdot
10^{-10}$. Thus, neglecting the lepton asymmetries (which is an
excellent approximation
unless they are much larger  than baryon asymmetry)
one can use the observations of primordial abundances to test if the
number of thermalized neutrinos is $N_\nu=3$
as predicted by the SM. 

In our analysis we fix the active-active oscillation parameters to
their experimental values, and given the sterile-active mixing
parameters we solve the $4 \times 4$ neutrino density matrix kinetic
equations and follow the relevant network of Boltzmann equations in
order to compute the $\he,$D abundances. At the end we convert the
computed abundances in an effective number of neutrinos. We consider
$\nnu \gtrsim 3.8$ disfavored by the experimental data \cite{He4},
while $\nnu \gtrsim 3.2$ could be tested if the experimental
determinations of the $\he,$D abundances improve.

{\bf Cosmic Microwave Background (CMB).} CMB anisotropies are sensitive to the energy content of the Universe at temperatures $T \sim \eV$, since it determines the pattern of fluctuations which is measured by WMAP (and other experiments). Neutrinos affect CMB in various ways \cite{Bashinsky:2003tk}. Global fits at the moment imply \cite{CMBfits}
$N_\nu^{\rm CMB} \approx 3 \pm 2$
somewhat depending on which priors and on which data are included in the fit.
Future data might start discriminating $3$ from $4$ neutrinos.

{\bf Large Scale Structure (LSS).} Neutrinos can be studied
  looking at the distribution of the galaxies because massive
  neutrinos move without interacting, making galaxies less
  clustered. Observations can constrain the neutrino relic density
  $\Omega_\nu h^2=\hbox{Tr} [m\cdot \rho]/(93.5\eV)$, where $m$ is the
  $4\times4$ neutrino mass matrix and $\rho$ is the $4\times4$
  neutrino density matrix. We approximate the present bound
  \cite{Spergel:2003cb} with $\Omega_\nu h^2< 0.01$. A $0.001$
  sensitivity in $\Omega_\nu h^2$ could be reached in the future.

  \begin{figure}[t]
    \centering
    \includegraphics[width=6cm]{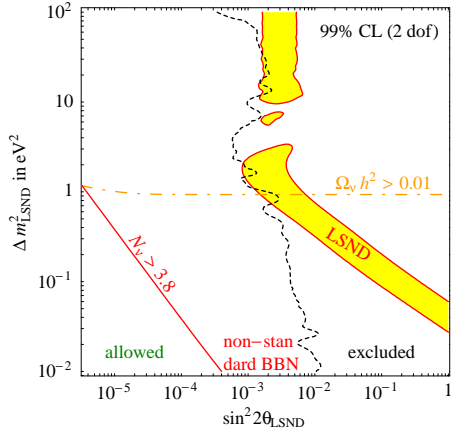}
    \caption{{\bf The LSND anomaly interpreted as oscillations of 3+1 neutrinos}.
Shaded region: suggested at 99\% C.L.\ by LSND.
Black dotted line: 99\% C.L.\ excluded from other neutrino experiments
(mainly Karmen, Bugey, SK, CDHS).
Continuous red line: $N_\nu = 3.8$ thermalized neutrinos.
Dot-dashed orange line: $\Omega_\nu h^2 =0.01$.}
    \label{fig:LSND}
  \end{figure}

For the technical details see \cite{Cirelli:2004cz}. The results are
shown in Fig. \ref{fig:all}. The relevant line is the red one, which
corresponds either to a number of effective neutrinos for BBN equal to
$3.8$ or to a neutrino relic density equal to $0.01$, which are both
excluded in standard cosmology. The bound from BBN dominates at lower
$\delta m^2$, while the bound from LSS is stronger at higher $\delta
m^2$ and small mixing angles.  For the $\nu_1/\nu_s$ mixing the bound
is stronger then for the $\nu_{2,3}/\nu_s$ mixing since $\nu_1$ has
the maximal $\nue$ component (see Fig. \ref{fig:spettrias}). BBN is
more sensible to $\nue$ since it enters directly in the weak reactions
which interconvert neutrons into protons. For the mixing with flavor
eigenstates the effect does not vanish for $\Delta m^2 \to 0$ since
the sterile component is present in all the mass eigenstates (see
Fig. \ref{fig:spettrias}a).

We also analyzed the impact of the LSND anomaly on cosmology.
 Extending previous analysis to the full 4 neutrino context, we
 show our results in Fig. \ref{fig:LSND}. We can conclude that
  the region which explains the LSND anomaly through a sterile neutrino
  is strongly disfavored by Standard Cosmology (BBN primarily) since
  it would correspond to a number $N_\nu =4$ of thermalized neutrinos which seams to be disfavored by
  $\he$ data.

\section{Sterile neutrinos in supernov\ae}
\label{sec:SN}

Supernov\ae{} arise from the gravitational collapse of stars. The collapse begins when the iron core of the star reaches the Chandrasekhar limit. The collapse abruptly stops when nuclear densities are reached: the falling material bounces on the surface of the inner core and turns the implosion of the core in an explosion of the outer layers. Although there is not yet a definite theory of supernova explosion a few key features are quite robust.  The beta reactions effectively act as a continuous pumping of energy and lepton number from the core matter (which gets neutronized) into the neutrinos, that carry them away and will give rise, at the end of the game, to a neutron star. Independently on the details which give rise to the supernova explosion, about 99 \% of the gravitational binding energy of the progenitor star (typically $10^{53}$ erg) is released through neutrino emission in a few tens of seconds. Despite the low statistics, observations  from SN1987A \cite{Hirata:1987hu} show an agreement between these estimates and the observed neutrino flux duration and intensity. If one adds channels where the active neutrinos can escape (sterile neutrinos in this case, but also for example extra dimensions \cite{Cacciapaglia:2002qr}) one has to handle with the so called ``energy constraint'': the total gravitational binding energy of the progenitor star is pretty well estimated and cannot be drastically overcome. All these considerations allow to use supernov\ae{} to put constraints on the mixing between active and sterile neutrinos. Other bounds could be considered. For example in the neutrino-driven picture of supernova explosion the active neutrinos flux is essential in order to provide the necessary energy to the outer material to successfully explode. However the ``energy constraint'' seems significantly more robust and it is the only one we use.

\begin{figure}[t!]
  \centering
  \includegraphics[width=15.3cm]{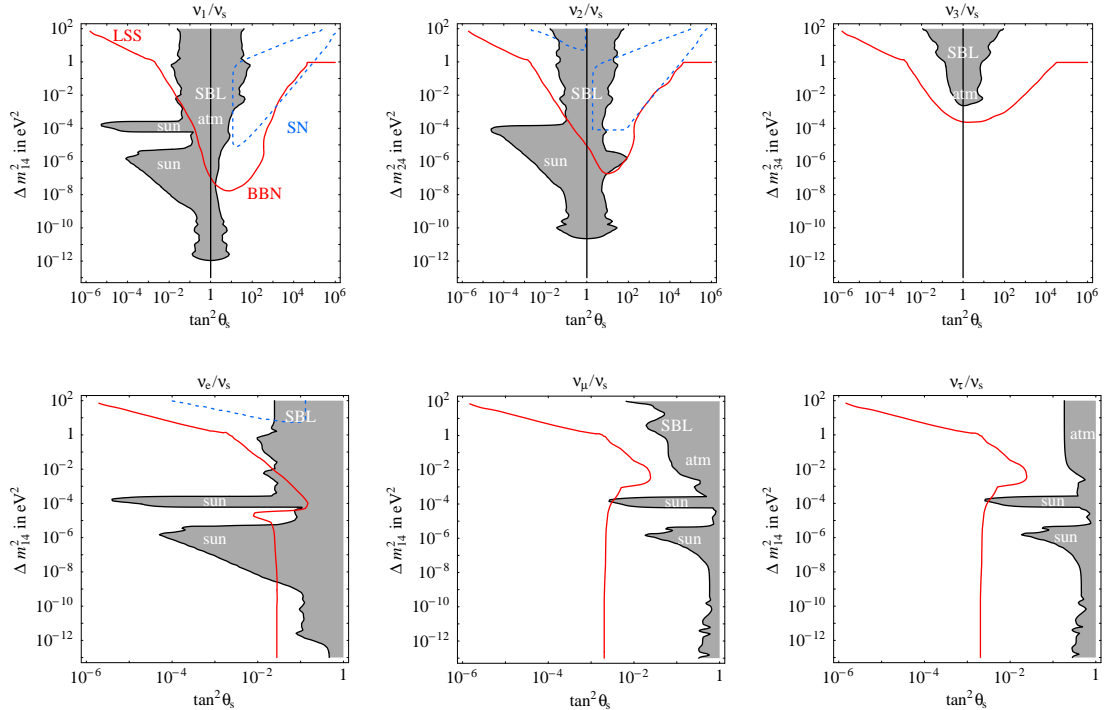}
  \caption{\label{fig:all} Summary of sterile neutrino effects. The dashed region is excluded at $99\%$ C.L.\ (2 dof)
by solar or atmospheric or reactor or short base-line experiments. The
red continuous line corresponds either to $3.8$  neutrinos
for BBN (for lower $\delta m^2$) or to a neutrino relic density
$\Omega_\nu = 0.01$ (for higher $\delta m^2$), both excluded by
standard cosmology. The dashed blue line corresponds to
  a suppression of the SN1987A $\bar\nu_e$ rate by more than $70\%$. }
\end{figure}

We compute the reduction of the flux of electron antineutrinos (which
have been measured by \cite{Hirata:1987hu}) with respect to the
standard case (no sterile neutrinos and active neutrinos mixed with
solar and atmospheric mixing parameters). The techniques to be used are
a generalization of the solar ones, with a significant complication
due to the fact that in a SN the matter density grows from zero up to
the nuclear densities, making matter effects very strong. A crucial
point is the matter density and the electron fraction profiles in the mantle of
the star. We used the results of \cite{Thompson:2002mw}. The results
are shown in Fig. \ref{fig:all} where we plotted the region which
produces a suppression of 70\% or more of the SN1987A $\nueb$ rate (dashed
blue contour). If a new supernova like SN1987A is observed in the future much
more statistics will be available. Furthermore, the theoretical
understanding of the supernov\ae{} explosion mechanism could
significantly improve. Then, supernov\ae{} are a promising tool to
test the physics of sterile neutrinos.

\section{Conclusions}

Neutrino physics is maybe the best field where the interplay between
experiments and cosmology can be seen. We reviewed the effects
of the mixing between active and sterile neutrinos taking into account
the established active-active mixing. We analyzed effects on solar, atmospheric, reactor and beam
experiments, cosmology
(BBN, CMB and LSS) and supernov\ae. We found no statistically significant evidence of sterile neutrinos and
identified the still allowed region of parameter space. We also
outlined the future promising signals: sub-MeV solar experiments, more
precise determinations of primordial abundances, new data for CMB, new
data and improvement of theoretical understanding of supernov\ae.

\section*{Acknowledgments}

I warmly thank Marco Cirelli, Alessandro Strumia and Francesco Vissani
for precious collaboration. It is also a pleasure to thank the
Organizing Committee of the XXXIXth Rencontres de Moriond.

\end{document}